%% file: main.tex
\documentclass[compsoc, conference, a4paper, 10pt, times]{IEEEtran}

\usepackage{cite}
\usepackage{amsmath,amssymb,amsfonts}
\usepackage{algorithm,algorithmic}  %% typset algorithms

\usepackage{graphicx}
\usepackage{textcomp}
\usepackage{xcolor}
\usepackage{booktabs}
\usepackage[hidelinks]{hyperref}

\usepackage{amsthm}
\usepackage{textgreek}
\usepackage{xspace}
\usepackage{centernot}

\usepackage{newunicodechar}
\newunicodechar{Π}{\ifmmode\Pi\else\textPi\fi}
\newunicodechar{⫫}{\mathrel{\perp\!\!\!\perp}}

\usepackage [autostyle, english = american]{csquotes}
\MakeOuterQuote{"}
\usepackage{graphicx}

\renewcommand{\paragraph}[1]{\vspace*{2pt}\noindent\textbf{#1}}

\newcommand{\langname}{\textsc{\textbf{.}CHO}\xspace}
\newcommand{\toolname}{\textsc{DT-SIM}\xspace}

\title{\toolname: Property-Based Testing for MPC Security}

\author{\IEEEauthorblockN{Mako Bates}
\IEEEauthorblockA{\textit{University of Vermont} \\
Burlington, Vermont, US \\
mako.bates@uvm.edu
}
\and
\IEEEauthorblockN{Joseph P. Near}
\IEEEauthorblockA{\textit{University of Vermont} \\
Burlington, Vermont, US \\
jnear@uvm.edu
}
}

\begin{document}

\newtheorem{definition}{Definition}

\ExplSyntaxOn    %% https://tex.stackexchange.com/a/639374/244649
\newcommand { \eg }
{ \textit{e.g.}
  \peek_meaning_ignore_spaces:NTF .
    { \skip_horizontal:n { -.3ex } \use_none:n }
    { \peek_meaning_ignore_spaces:NF , { \skip_horizontal:n { -.3ex } } }
}
\newcommand { \ie }
{ \textit{i.e.}
  \peek_meaning_ignore_spaces:NTF .
    { \skip_horizontal:n { -.3ex } \use_none:n }
    { \peek_meaning_ignore_spaces:NF , { \skip_horizontal:n { -.3ex } } }
}
\ExplSyntaxOff

\newcommand{\termOfArt}[1]{\textbf{#1}}

\maketitle
\thispagestyle{plain}
\pagestyle{plain}

\begin{abstract}
  Formal methods for guaranteeing that a protocol satisfies a cryptographic security definition have advanced substantially,
  but such methods are still labor intensive and the need remains for an automated tool that can positively identify an insecure protocol.
  In this work, we demonstrate that property-based testing, \textit{"run it a bunch of times and see if it breaks"},
  is effective for detecting security bugs in secure protocols.
  We specifically target Secure Multi-Party Computation (MPC), because formal methods targeting this security definition for bit-model implementations
  are particularly difficult.
  Using results from the literature for Probabilistic Programming Languages and statistical inference, we devise a test that can
  detect various flaws in a bit-level implementation of an MPC protocol.
  The test is grey-box; it requires only transcripts of randomness consumed by the protocol and of the inputs, outputs, and messages.
  It successfully detects several different mistakes and biases introduced into two different implementations of the classic GMW protocol.
  Applied to hundreds of randomly generated protocols, it identifies nearly all of them as insecure.
  We also include an analysis of the parameters of the test, and discussion of what makes detection of MPC (in)security difficult.
\end{abstract}

\section{Introduction}

Secure multiparty computation (MPC) protocols~\cite{evans2018pragmatic} allow a group of mutually distrusting parties to compute a shared function over distributed data without revealing that data. The security of MPC protocols is often expressed as a negation,
\eg{} secret information should \textit{never} be derivable from an adversary's observations of the system.
Unfortunately, testing generally provides limited evidence that the system is actually secure in the intended ways; formal proofs of security are effective, but are tedious to construct and require significant expertise. As a result:
\begin{itemize}
\item Most implementations of MPC protocols are not formally verified.
\item Formal verification tools are challenging to use, and engineers can waste significant time trying to prove properties of insecure implementations.
\item Even for formally-verified software, we would like to have sanity checks
  that protocol implementations actually have the security properties their constituent source-code claims to be proven to have.
\end{itemize}

We present \toolname, an \emph{automatic}, \emph{property-based testing} tool for MPC security. \toolname works by running the protocol thousands of times, then using a statistical independence test to detect security bugs. Our approach is motivated by the success of property-based testing (via tools like \textsc{QuickCheck}) as an approach for finding bugs in software~\cite{fink1997property, claessen2000quickcheck, paraskevopoulou2015foundational}. Unlike previous applications of property-based testing, however, \toolname attempts to falsify a \emph{probabilistic hyperproperty} of the target protocol---a much more challenging setting.

\toolname works by defining security in terms of conditional independence, then performing a statistical test which attempts to falsify this conditional independence property. In MPC protocols, the inputs of the honest parties should be independent of the \emph{views} (\ie{} the observed message values) of the corrupt parties, conditioned on the protocol's output. Conditioning on the protocol's output is vital for the security definition, since it allows the protocol to \emph{intentionally} leak some information about the inputs via the output---as nearly all MPC protocols do. Manual security proofs for MPC protocols demonstrate this conditional independence property by constructing a \emph{simulator} that ``fakes'' the corrupt parties' views using only the protocol's output.

This formulation of security as conditional independence puts MPC security out of reach of all automated verification tools (and many manual ones). While automated tools have been used to verify security in cryptographic protocols~\cite{gancher2023owl, barthe2019probabilistic, darais2019language, fournet2011information}  (\eg{} key exchange protocols), these tools are limited to protocols whose \emph{outputs reveal nothing} to the adversary. They are capable of showing independence, but not the kind of \emph{conditional} independence required for MPC security. Recent manual approaches could be used to verify MPC security~\cite{li2023lilac, gancher2023core, haagh2018computer}, but require manual proof.

\toolname is designed to fill the same role as tools like \textsc{QuickCheck}: it can be used to quickly and automatically detect many classes of bugs in MPC protocol implementations, without investing the user's time in a manual security proof.
Using \toolname requires only that key runtime values can be captured during program execution;
it's agnostic to the language and many other details about the subject program.
We have implemented \toolname and tested it empirically on several real-world MPC protocols and on hundreds of randomly-generated protocols; our experimental results suggest that \toolname is capable of scaling to realistic protocols and detecting many security bugs, despite the challenging setting.

\paragraph{Contributions.}
In summary, our contributions are:
\begin{itemize}
\item We show how to define MPC security in terms of conditional independence, enabling the use of independence testing tools for checking MPC security.
\item We introduce \toolname, the first property-based testing tool for MPC security.
\item We conduct an empirical evaluation of \toolname, demonstrating its ability to scale to realistic MPC protocols and detect security bugs.
\end{itemize}

\section{Background}

\subsection{MPC security}

Secure Multi-Party Computation (MPC or SMPC, depending on the text)
is a cryptographic process that allows some group of participating parties or machines
to compute a function that depends on all of their private inputs,
without any of them learning anything except the final result\footnote{
    It is possible for each party to have a distinct final result.
}.
Implementations that work on arbitrary functions (representable as circuit computations in some field)
have existed since 1986~\cite{yao1986generate}, % https://ieeexplore.ieee.org/document/4568207
but their use by the public is limited by their performance.
Some performance limitations are fundamental to the problem
(roughly, every branch of the computation must be executed so that the control-flow does not itself leak information),
but techniques have been found to improve performance somewhat.

The traditional expression of MPC assumes a pre-determined function F, the \emph{functionality},
mapping the tuple of all participants' input vectors to the respective tuple of output vectors
(or to distributions of outputs, if F is non-deterministic).
In an \termOfArt{ideal world}, a trusted third party (the \termOfArt{ideal functionality}) would accept input messages
from all other parties, compute F, and then send the parties their respective outputs.
A \termOfArt{real world} protocol Π is said to \textit{correctly} implement F if, for all possible inputs,
Π results in all parties getting the same outputs as they would have in the ideal world.
The \textit{security} of Π is expressed in terms of a particular threat model and the participants' "\termOfArt{views}".

\textbf{Passive security}
supposes a computationally-bounded adversary
that can see everything happening inside a fixed subset of the participating machines.
This adversary cannot cause the parties to deviate from the protocol Π in any way;
we are interested in whether it can learn anything (beyond the outputs of the "corrupted" machines)
by observing the execution of Π.
As an example, this notion of security might be of interest if Alice trusted Bob to participate
in a distributed computation with her, but did not trust him to properly store or destroy the log-files
generated.
The \termOfArt{ideal world view}, $\mathsf{Ideal}_P(F,\vec{x})$, of a party P is what it sees when running F in the ideal world:
just its inputs $\vec{x}_P$ and its outputs $\vec{y}_P$.
The \termOfArt{real world view}, $\mathsf{Real}_P(Π, \vec{x})$, of a party P is everything it sees while executing its part of Π in the real world:
its inputs, outputs, all messages it receives, and all the bits of random tape it uses.
Both these concepts extend naturally to lists of parties.

\begin{definition}[Passive MPC Security]\label{def:passive-mpc}
    A protocol Π,
    involving participating parties P,
    that correctly implements functionality F,
    is secure against a passive adversary controlling C$\subset$P iff
    there exists a poly-time-computable function $\mathtt{SIM}$ s.t.
    for all combinations of inputs $\vec{x}$
    $$\begin{aligned}
        \mathsf{Real}_C(Π, \vec{x}) &\cong \mathtt{SIM}(\mathsf{Ideal}_C(F,\vec{x}))
    \end{aligned}$$
    where $\cong$ denotes distributional equivelence or indistinguishabilty.
\end{definition}

Definition~\ref{def:passive-mpc} represents security by a transitive argument:
Since the adversary in the ideal world can compute all the data the real-world adversary starts with,
anything the real-world adversary can compute the ideal-world adversary can also compute!
The hypothetical function $\mathtt{SIM}$ is called a "simulator";
it simulates in the ideal world what the real-world adversary would see.
Often MPC security proofs will give an explicit simulator function.

In practice, security is usually quantified over all possible corruptions $C$
where $\lvert C \rvert \leq \frac{\lvert P \rvert}{r}$ for some $r$, but this is not fundamental and we only consider a single $C$ at a time.

Definition~\ref{def:passive-mpc} is tight, in the sense that attempting to relax it invariable either
allows a "secure" protocol to do something bad, or requires introducing restrictions
on the kinds of functionality to which the definition may be applied\footnote{
    The idea that a secure protocol \textit{can't} do anything bad assumes that computing the functionality is good to begin with.
}.
That said, some simplification is possible for our purposes:
Since the (passive) corrupted parties are assumed to follow Π faithfully, we are fine to limit ourselves to
\termOfArt{choreographic} protocols in which all messages sent are expected and all messages expected are received.
In this context, and because the only source of non-determinism is the parties' random tapes
(which are captured in the real-world views),
we can safely ignore any messages sent \textit{within} $C$.
Then we can collapse all the corrupted parties $C$ into a single party "C",
and similarly all the other parties into a single party "H" (for "honest")\footnote{
    We don't always do this, it's not required for \toolname to work.
}.
Finally, in this work we assume that protocols implement their functionalities \textit{correctly};
the corresponding F is implicit.
In other words, we define F to be whatever mapping results from the execution of Π.
This present work describes a system for property-based testing of MPC \textit{security};
a corresponding test of \textit{correctness} is trivial for any pre-defined F.

\textbf{Active security}
(sometimes "malicious security")
supposes a computationally bounded adversary that can cause corrupted machines to take arbitrary actions:
breaking the choreography,
forging messages,
lying about computation results,
and/or playing along with the protocol for any period of time.
In contrast to passive adversaries, which are weaker than the threat models most real programmers worry about,
an active adversary subsumes any threat model (including adaptive corruption, depending how it's framed).
The corresponding security definition is a bit more involved, and we omit it for brevity.
Since we're not directly testing for active security in this work,
all else that needs to be said about it is that
any protocol which is not \termOfArt{passive secure} is automatically not \termOfArt{active secure}.

\paragraph{Vacuousness.}\footnote{
    This terminology is new; we needed it to talk though problems we encountered in during this project.
}
Certain edge cases are helpful to consider, as they map out the edges of the problem of testing from MPC security.
In the first case, Π itself may be vacuous.
A \termOfArt{vacuous protocol}, one in which no information (random bits don't count), is communicated is automatically secure,
although of course it can only correctly implement trivial functionalities.
On the other hand, if the ideal-world adversary can efficiently deduce a unique possible honest input upon observing the corrupt output,
then the protocol is \termOfArt{vacuously secure}: \textit{any} correct implementation of that functionality is secure regardless of what's communicated.
These extremes mark the ends of a spectrum; most protocols are somewhere in the middle,
where learning the output focuses the corrupt party's distribution-estimate of the honest secrets but still leaves some uncertainty.
On the other hand, this spectrum doesn't capture the assumptions about the adversary's computational limits;
if inverting the functionality F is NP-hard, then the protocol is \textit{not} vacuously secure, even if F is a bijection.

\paragraph{Non-interference.}
While MPC can be expressed as a kind of probabilistic non-interference property~\cite{haagh2018computer, almeida2018enforcing}, this is only somewhat useful.
The non-interference can only be meant within the space of possible program traces \textit{posterior to the output}.
Trying to avoid such subtleties in the non-interference property by providing a correctly constrained declassification operator
does not avoid this problem: constraints on declassification would need to recapitulate the original question of security
and examine the entire rest of the protocol as context.

\subsection{Inference in PPLs}

Since MPC security can be framed as a test of conditional independence
(Section~\ref{sec:algorithm}),
\termOfArt{probabilistic programming languages} (PPLs) with efficient inference and independence tests could hypothetically be used to verify MPC security.
A PPL is a language that provides stochastic non-deterministic behavior as a language primitive
and provides meta-theory or auxiliary tooling to reason about the resulting distributions of program behavior.
In particular, PPLs commonly have semantics defined in terms of transformations on distributions,
and are designed to ensure that the resulting distributions are amenable to specified techniques for efficient
inference and sampling. Since even representing a fully-general multivariate distribution takes exponential space with respect to the number of variables,
PPLs typically restrict the programs and/or distributions they can handle.

For application to the subject of MPC protocols, a PPL would need to handle the two different kinds of uncertainty facing a hypothetical adversary:
(1) Random values generated (by any party) at runtime have known distributions\footnote{
    \eg{} in Python \texttt{secrets.SystemRandom().random()}
    will sample at runtime from the best floating-point approximation the machine can provide
    of the uniform distribution on $\mathbb{R}_{(0,1]}$
}, and
(2) The secret inputs of the honest parties are not stochastic but the adversary can be asked to provide a distribution over
possible secrets representing their prior belief so that the secret inputs can also be modeled as random variables.

Hypothetically, supposing a protocol in a PPL providing a multivariate distribution over an appropriate subset of all runtime values,
if it were possible to efficiently infer and sample from the real-world views
(or other variables sufficient to deterministically compute a view)
\textit{posterior to} the corrupt outputs (and corrupt inputs),
then that would suffice as a simulator
\textit{if and only if} the inferred distribution were independent of the adversary's choice of prior distribution for the honest secrets.

A direct test of MPC security via a PPL would require \emph{both}
(A) a language meta-theory guarantee that such posterior inference and sampling would be efficient, \emph{and}
(B) a test or guarantee of independence between the posterior view distribution and the exponentially-many parameters
of the prior distribution of honest secrets\footnote{
    Presumably such process would need to leave those parameters implicit.
    Attempting to represent the adversary's prior beliefs more compactly would not produce a valid proof or test of security.
}, as well as
(C) the ability to quantify (A) and (B) over all possible corrupt secrets or public inputs, presumably by handling them symbolically.
While some PPLs can statically verify independence properties~\cite{gorinova2021conditional}, none can verify the kind of \emph{conditional} independence required for MPC security. Lilac~\cite{li2023lilac} provides a separation logic for reasoning about conditional independence, but is not automated.

\subsection{Property-based testing}

In industry the normal way to see if a program satisfies some desired property is to run it and see what it does.
Sometimes this is done with a select handful of inputs or configurations,
but non-deterministic programs are better evaluated with \termOfArt{property-based testing}~\cite{fink1997property, claessen2000quickcheck, paraskevopoulou2015foundational},
in which the program is run with many arbitrary (but not necessarily random) configurations.
Property-based testing is a popular paradigm for testing code in industry;
the classic Haskell framework \textsc{QuickCheck}~\cite{claessen2000quickcheck}
is still in use and has been ported to various other languages.

At face value, property-based testing doesn't extend well to hyper-properties,
which can only be assessed by comparing multiple program traces.
In practice sometimes one can express a hyper-property as a property of multiple duplicate programs,
but the sensitivity of the test quickly decays and the logistics of setting up the test quickly grow.
Furthermore, when the property to be tested is over distributions of traces,
the techniques used by good property-based testing frameworks to efficiently find interesting cases are not appropriate.

\section{\toolname: a Statistical Test for Security}

This section describes \toolname, our approach for using statistical independence testing to find security bugs in MPC protocols. The approach is based on the \emph{fast conditional independence test} proposed by Chalupka et al.~\cite{chalupka2018fast}.

\subsection{Conditional Independence Testing by Classification}

Chalupka et al.~\cite{chalupka2018fast} present a high-dimensional conditional independence test based on the observation that if $P(X | Y, Z) = P(X | Y)$ (i.e. $X$ and $Z$ are conditionally independent given $Y$), then $Y$ and $Z$ together will not be more useful for predicting $X$ than $Y$ alone. They leverage this observation to design a test based on training machine learning classifiers: one classifier predicts $X$ from both $Y$ and $Z$, while the other predicts $X$ from $Y$ alone. If the first classifier yields better test error than the second, then the conditional independence property likely does not hold.

Chalupka et al. use decision trees (D-Trees) as their ML classifiers.
Because the process requires training many copies of the model, a light-weight model like a D-Tree is preferable;
we also use D-Trees for \toolname.
In principal, \textit{any} ML classifier would be ok to use; the choice is partly an engineering question
and partly a judgment call about what kinds of systems will be best able to discern patterns in the views of the subject protocol.

\subsection{The \toolname Algorithm}\label{sec:algorithm}

If a simulator $\mathtt{SIM}$ exists for a protocol Π implementing functionality F,
and we suppose fixed corrupt inputs $\vec{x}_C$,
then for any observed $\vec{y}_C$ the values yielded by $\mathtt{SIM}(\vec{x}_C, \vec{y}_C)$
will be drawn from the same distribution regardless which of the consistent honest inputs $\vec{x}_H$ were actually used.
This is conditional independence, and since we assume the simulator is correct (Π is secure),
we have that independence between the honest secrets and the corrupt real-world views conditional on the corrupt ideal-world views
is a requisite feature of MPC security.
In other words

$$
\left( \vec{y}_H \centernot{⫫} \mathsf{Real}_C(Π, \vec{x}) \mid \mathsf{Ideal}_C(F,\vec{x}) \right)
\rightarrow
\nexists \; \mathtt{SIM}
$$

We use the techniques from Chalupka et al.~\cite{chalupka2018fast} to build \toolname,
which directly tests for the above conditional dependence without any static analysis of the subject protocol.

To actually test a protocol, \toolname uses two identical machine learning (ML) tools to predict the honest secrets based on the corrupt views.
One of the two models has access to the real-world views, and the other only has access to the ideal-world views.
Generating the data for these tests requires running the protocol many times.
The process is enumerated in Algorithm 1.
Note that the data must be understood as vectors/matrices/tensors of bits;
aggregating values into larger structures such as bytes or integers would at best require careful modifications to line 8 in Algorithm 1.

\begin{algorithm}  %% https://tex.stackexchange.com/a/219820/244649
  \caption{Test for MPC insecurity}
  \begin{algorithmic}[1]
    \STATE Set a chosen number of iterations \texttt{iters},
           number of rows of data on which to train each model \texttt{trainN},
           and number of rows of data on which to test each model \texttt{testN}.
  \STATE Choose $\alpha$, your probability of randomly flagging a secure protocol as insecure.
    \FOR {\texttt{i$\leftarrow$[1..iters]},
          calculate scores for the real and ideal models:  \\}
      \STATE Run the protocol \texttt{trainN} times,
             with uniformly random inputs,
             collecting the corrupt views and honest secrets.
      \STATE Train the first ML model on the collected data.
             The real-world views are the input features, and the honest secrets are the labels.
      \STATE Train the second ML model the same way, but occlude view data to only reveal the ideal-world view.
      \STATE Run the protocol \texttt{testN} more times, again collecting the relevant data.
      \STATE The score for each model is the $L^1$ distance between the testing labels and the respective model outputs
             (\ie the number of bits wrong when run on the \texttt{testN} rows of test data).
    \ENDFOR
    \STATE Using the \texttt{iters} samples of real-world and ideal-world scores,
           calculate the $p$-value for the hypothesis that
           the distribution of real-world scores is better on average than the ideal-world distribution.
    \IF {$p \leq \alpha$} \RETURN \texttt{INSECURE} \ELSE \RETURN \texttt{MAYBE SECURE} \ENDIF
  \end{algorithmic}
\end{algorithm}

\subsection{Interpreting Test Results}

In principal the only aspect of MPC (in)security that conditional (in)dependence fails to capture
is the constraint on a hypothetical adversary's computational power.
In other words, a protocol may truly have the appropriate conditional independence,
but still be insecure because access to the real-world views allows an adversary to compute in polynomial-time
something that would require super-polynomial-time in the ideal world.
We believe this to be a much smaller concern in practice than the statistical limitations of independence testing.

The usual trade-offs of significance testing apply.
Setting $\alpha$ too low will cause programs whose insecure-ness is only marginally discernible to the D-Trees
to pass the test.
Setting $\alpha$ higher will compromise one's certainty that a failing program truly is insecure.
There are three reasons why this is not a serious problem with \toolname, each worth considering as its own subtlety.

First, if \toolname returns \texttt{INSECURE},
then there's nothing wrong with running it again to see if you were just unlucky.
This assumes that you have the resources (\eg{} compute time) to do so, which you might not.
Since more powerful tests are better able to detect problems,
if your compute-budget is fixed you're better off spending it all on a single bigger test.

Second, we observe that the transition of an insecure program between "appears secure" and "appears insecure"
as the power of the test increases is sharp,
and as the size of the test continues to increase the $p$-values quickly drop below any reasonable $\alpha$.
This can be frustrating; if you're unsure of the security of a protocol
then there's always a chance a modestly-more-powerful test might be conclusive.
On the other hand, it means that at any reasonable fixed test-size,
most of the protocols \toolname \textit{can} detect insecure will be \textit{unambiguously} insecure.

The third such mitigating factor is also a problem on its own.
For a secure program the observed $p$-values
will typically \textit{not} be sampled from the uniform distribution on $\mathbb{R}_{(0,1]}$,
and $\alpha$ is actually an \textit{upper bound} on the likelyhood of falsely flagging a protocol as insecure.
Contrary to casual descriptions of significance testing, this is not a problem with the statistics;
it's a result of the inequality in the testing hypothesis.\footnote{
    Significance testing of an \textit{equality} based null-hypothesis would yield uniformly distributed $p$-values when the null-hypothesis is true.
}
Because the ML models have no knowlege of the importance or semantics of the bits they're trained on,
the (mostly random) message- and tape-data provided to the real-world models makes it harder for them
to learn from the input- and output-data they share with the ideal-world models.
In other words, the additional data on which the real-world models are trained is fundementally a hindrance to them,
and only becomes advantageous if the protocol is "insecure enough" to overcome this hidrance.
A protocol might be truely insecure and not have the relevant conditional independence,
but still appear secure because the advantage an adversary could gain at guessing the honest inputs
is smaller than the advantage the ideal-world ML models gain from their more focused data.
This issue could be overcome by setting \texttt{trainN} sufficiently high;
the ideal-world models' scores will saturate before the real-world models' do.

One might fix the above problem by padding the ideal-world views with random noise out to the same width as the real-world views.
In our experiments this introduced an even worse problem.
The additional data afforded to the real-world models is not fully random;
the whole view is self-consistent per the semantics of the protocol.
Therefore, the uniformly-random padding given to the ideal-world model will be "heavier" than what it's supposed to be balancing,
invalidating a key assumption of the whole test:
that any advantage of the real-world models comes from insecurity in the protocol.

\subsection{Implementation}

Our \toolname Python script consumes a stream of fixed-length bit vectors as a CSV.
Annotation of which columns are available in the real and ideal worlds is provided in the header.
The CSV is parsed and broken up as needed as a \texttt{pandas} dataframe.
The individual ML models are instances of \texttt{sklearn.tree.DecisionTreeClassifier},
and they're trained in parallel with \texttt{joblib.Parallel}.
The D-Tree scores are inflated by $+1 \cdot 10^{-10}$ to avoid division-by-zero in the score-ratio,
and the score distributions are compared by \texttt{scipy.stats.wilcoxon}.

Our implementation and the code to reproduce our experimental evaluation are available as open-source.\footnote{https://github.com/ShapeOfMatter/dt-sim}

\section{Evaluation}

We evaluate the test in terms of three standards of performance:
\begin{itemize}
\item \textbf{RQ1:} Is \toolname capable of detecting insecurity in novel protocols outside existing MPC paradigms?
\item \textbf{RQ2:} Is \toolname capable of detecting insecurity in implementations of real MPC protocols?
\item \textbf{RQ3:} Is \toolname scalable to realistic protocols in reasonable time?
\end{itemize}
We address \textbf{RQ1} in Section~\ref{sec:random-protocols}
and \textbf{RQ2} in Section~\ref{sec:case-study-protocols}.
In both cases we measure runtime and discuss implications for \textbf{RQ3}.
To facilitate all our experiments,
We built a custom DSL for representing protocols at the appropriate granularity and for easy integration with the test script
(Section~\ref{sec:cho-lang}).

\subsection{A Choreography Language}
\label{sec:cho-lang}

To facilitate analysis of the tool, we implemented a choreographic programming language, \langname,
with surrounding tooling to support large batches of parallel evaluations of a single program and piping of the generated data.
(In particular, the semantics are polymorphic within the Haskell type-class \texttt{Data.Bits.FiniteBits},
\eg{} sixty four evaluations can run in a single thread by applying the semantics to the \texttt{Word64} type.)
The design of \langname is driven by the contextual assumptions of MPC security.
In addition to bit-level computations and transmission of single bits between arbitrary parties,
each party has its own read-only tapes for secret inputs and randomness,
and its own tape for output bits.
\langname supports macros, but they are extremely limited.

\langname supports 1-of-N oblivious transfer as a primitive operation.
Oblivious transfer (OT)~\cite{kilian1988founding} is a 2-party protocol that allows one party ($R$, the \emph{receiver}) to select one out of several secret inputs provided by the other party ($S$, the \emph{sender}). In 1-out-of-2 OT, $S$ provides 2 secret inputs $x_0$ and $x_1$ and $R$ provides a secret \emph{selection bit} $b$. If $b=0$, $R$ receives $x_0$; if $b=1$, $R$ receives $x_1$. $R$ does not learn anything about the not-selected secret, and $S$ does not learn which secret $R$ selected. This idea can be extended to 1-out-of-$n$ OT by extending the selection bit $b$ to a $\log_2(n)$-bit string.
Oblivious transfer is a building block for many other protocols, including the GMW protocol described later.

An example \langname program is provided in Appendix~\ref{sec:example-cho}.

\subsection{Bug-Finding in Generated Protocols}
\label{sec:random-protocols}

The goal of this work is a tool that can detect mistakes in diverse MPC protocols.
We evaluate the generality of the approach by randomly generating hundreds of protocols and testing them with different test parameters.
We also use a batch of randomly generated protocols to asses the effects of the three test parameters
(\texttt{iters}, \texttt{trainN}, and \texttt{testN})
on the test's power and runtime.

Randomly-generated protocols are likely to be insecure, but the probability of generating a secure protocol is not zero.
In particular, some generator settings will be biased toward vacuous protocols in which nothing is communicated
or vacuously secure protocols in which the honest secrets are fully revealed by the ideal functionality.
% (Never mind that such protocols are useless; none of the protocols generated in this experiment are \textit{useful}.)
For reasonable generator settings we find that very few generated protocols
continue to appear secure as we increase the power of the test.

\paragraph{Experiment setup.}
In our first two experiments, we generate random protocols and evaluate \toolname's ability to detect insecurity in these protocols.
We randomly generate protocols according to the grammar of \langname.
The generator uses a specified size for each party's secret inputs, random tape, and outputs,
parameters for the length of the "body" of the program and the "max width" of each computation step,
and a handful of bias-parameters that affect the frequency of different operations.
We bias the generated protocols in favor of logical depth;
values (inputs, coin-flips, messages-received, values calculated) are likely
to be used on each new line in inverse proportion to the number of times they've already been used.
Finally, to focus our experiment on protocols for which there's any question of security,
\langname protocols that \toolname flags as insecure with very low power-settings
are filtered out of the stream of generated programs before we include them in our experiments.

We generated 500 programs with 500-line bodies (16-bit inputs and outputs, 48-bit random tapes)
and evaluate \toolname's ability to detect insecurity as \texttt{trainN} increased.
Within this experiment we locked \texttt{iters} at 100, and set \texttt{testN} as close to $\frac{\mathtt{trainN}}{4}$ as our setup allowed.
Each generated \langname file was tested with increasing values of \texttt{trainN} until the observed $p$-values fell below $1.25 \cdot 10^{-4}$.

To observe the relative role of the three test parameters, we generated 200 additional programs using the same settings
and ran \toolname against each of them with a grid of different parameters.

\begin{figure*}
  \centering
  \newcommand{\gsize}{.9\textwidth}
\begin{tabular}{c}
    \hline\hline
    \includegraphics[width=\gsize]{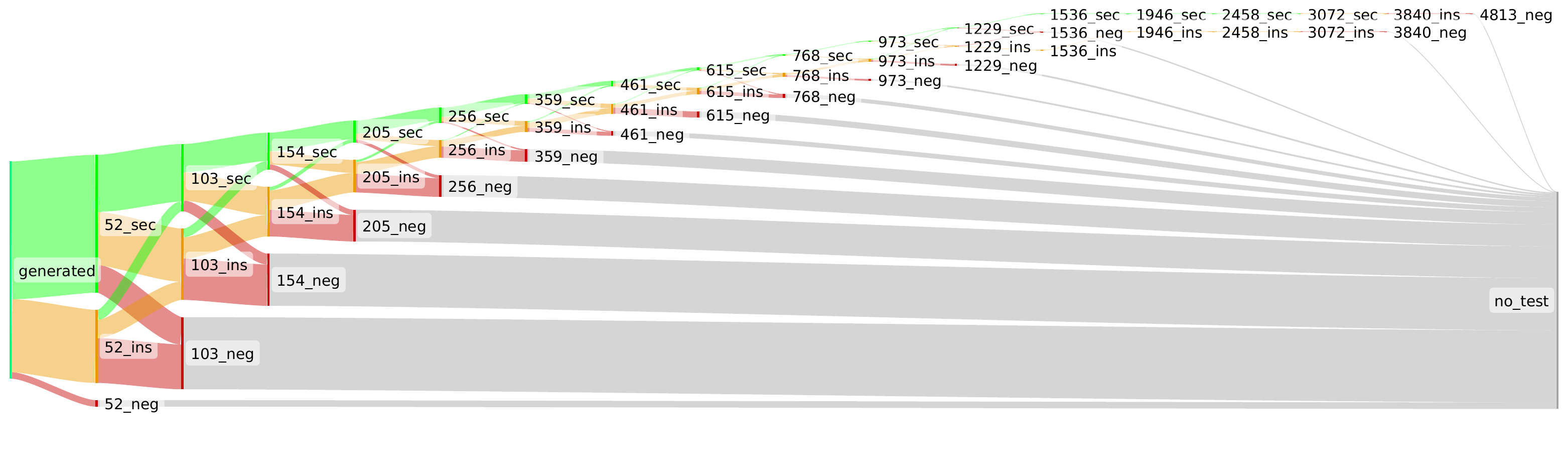} \\
    \hline
    \hline
\end{tabular}
\caption{Experimental results: performance on random protocols under increasing test power.}
\label{fig:sankey}
\end{figure*}

\begin{figure}
  \centering
  \newcommand{\gsize}{.38\textwidth}
\begin{tabular}{c}
    \hline\hline
    \includegraphics[width=\gsize]{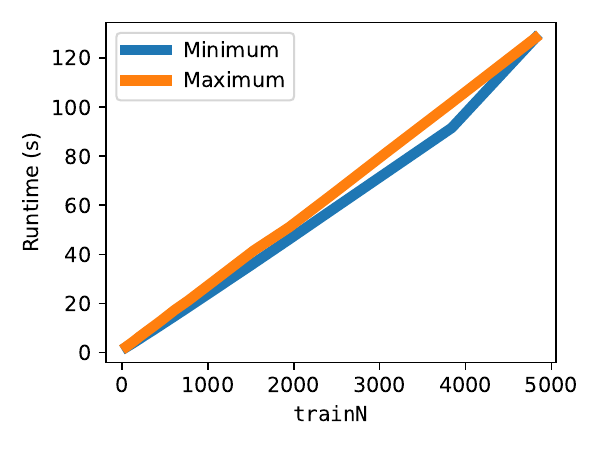} \\
    \hline
    \hline
\end{tabular}
\caption{Experimental results: runtime with increasing test power.}
\label{fig:linear-time}
\end{figure}

\begin{figure*}
  \centering
  \newcommand{\gsize}{.45\textwidth}
\begin{tabular}{c | c c}
    \hline\hline
    \rotatebox{90}{\phantom{hellohello}(a). Detection scores (summed)}
  & \includegraphics[width=\gsize]{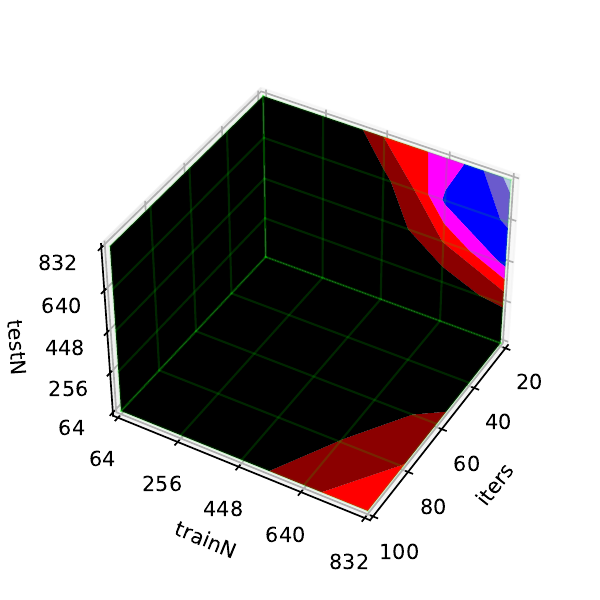}
  & \includegraphics[width=\gsize]{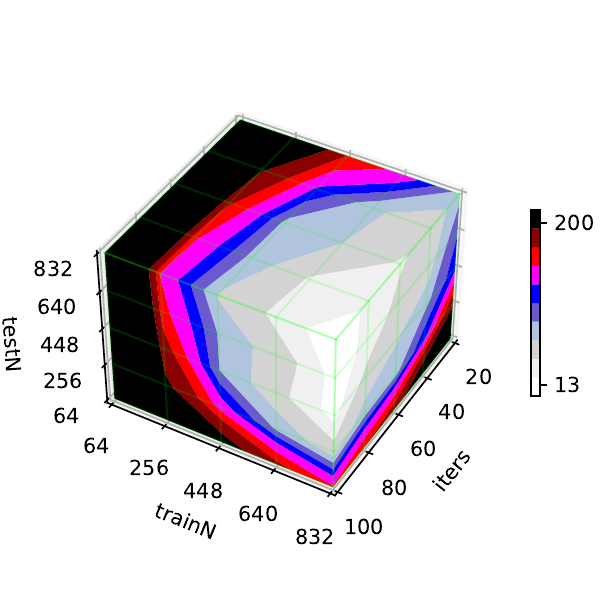} \\
    \hline
    \rotatebox{90}{\phantom{hellohellohello}(b). Mean runtime (s)}
  & \includegraphics[width=\gsize]{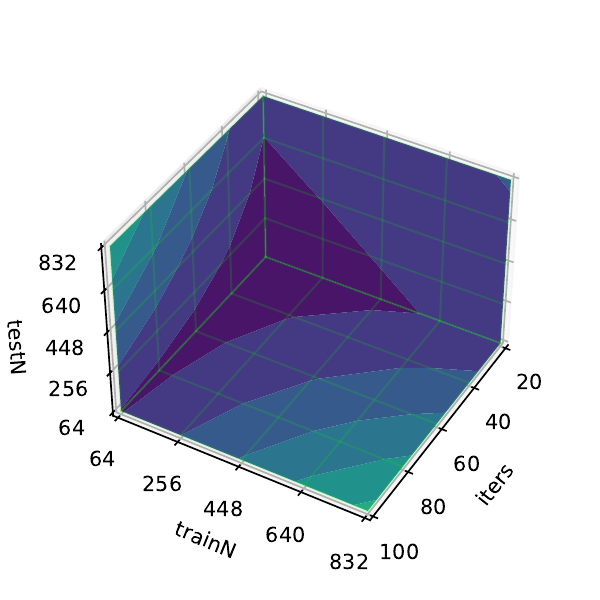}
  & \includegraphics[width=\gsize]{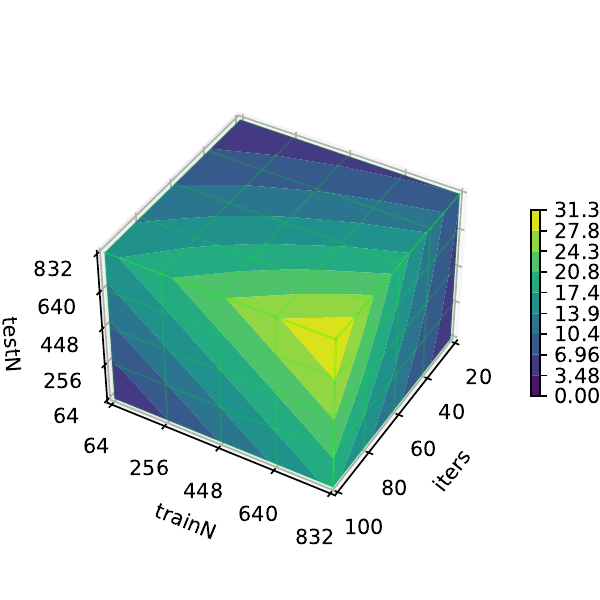} \\
    \hline
    \hline
\end{tabular}
\caption{Experimental results: trade-offs in power and runtime under different test parameters.}
\label{fig:cube-search}
\end{figure*}

\paragraph{Results.}
Figure~\ref{fig:sankey} shows that, while some programs appear secure at weaker test powers,
all the programs generated with these settings were identified as insecure with $\texttt{iters}=100$ and $\texttt{trainN}\leq4000$.
Specifically, this Sankey plot shows 500 randomly generated \langname files moving between buckets labeled
"\texttt{\#\_sec}" (for secure),
"\texttt{\#\_ins}" (for insecure at $\alpha=0.05$),
and "\texttt{\#\_neg}" (for "negligible", insecure at $\alpha=1.25 \cdot 10^{-4}$)
as they are tested by \toolname with increasing \texttt{trainN}.
Because there's always a chance (at most $\alpha$) of a secure protocol being flagged insecure,
\langname{}s sometimes move back to "secure" from "insecure"; we continue testing them until they reach "negligible".
Figure~\ref{fig:linear-time} shows time data from this same experiment;
because fewer programs were subjected to the larger tests, the minimum and maximum runtimes converge at the end.
More importantly, the runtime scales linearly with the amount of data used in the test.

Figure~\ref{fig:cube-search} shows the individual effects of \texttt{iters}, \texttt{trainN}, and \texttt{testN} on \toolname's power and runtime.
Each 3D heatmap shows aggregate data for 200 randomly generated \toolname files.
Because runtime is mostly driven by the amount of data consumed ($\texttt{iters} \cdot (\texttt{trainN} + \texttt{testN})$),
all three affect runtime linearly.
To maximize the granularity of the test-power analysis, we calculate a score for each combination of parameters,
where apparently-secure programs count for 2 points,
programs insecure at $\alpha=0.05$ count for 1 point,
and programs insecure at $\alpha=0.01$ count for zero.
Lower scores are "better" because we know all the programs turn out to be actually insecure.
We observe that scaling all three parameters is desirable,
but \texttt{trainN} should generally be larger than \texttt{testN},
and \texttt{iters} can be much smaller.

\subsection{Bug-Finding in Real-World Protocols}
\label{sec:case-study-protocols}

Our second set of experiments evaluates \toolname's ability to detect security bugs in real-world protocols. We implement two MPC frameworks for evaluating circuits, instantiate them with three different functions to obtain specialized protocols, systematically introduce bugs into these protocols, and evaluate \toolname's ability to detect the bugs. We summarize the implemented protocols in Section~\ref{sec:protocols-evaluated}, describe the experiment in Section~\ref{sec:e2-experiment-setup}, and present the results in Section~\ref{sec:e2_results}.

\subsubsection{Protocols Evaluated}
\label{sec:protocols-evaluated}

For our evaluation, we consider two MPC frameworks for evaluating circuits, and instantiate those frameworks with three different functions (represented as circuits) to construct secure protocols. The frameworks and functions are listed in Table~\ref{tbl:protocols} and described below.
All six protocols are designed for two parties, with the addition of a third ``dealer'' party to produce correlated randomness in several of the protocols.

\begin{table}
  \centering
  \renewcommand*{\arraystretch}{1.2}
  \begin{tabular}{|l l|}
    \hline
    \textbf{MPC Frameworks} & \textbf{Approach} \\
    \hline
    GMW~\cite{goldreich2019play, goldreich2009foundations} & Oblivious Transfer\\
    MPC w/ Beaver Triples~\cite{beaver1992efficient} & Beaver Triples\\[4pt]
    \hline
    \textbf{Instantiated Functions} & \textbf{Invertibility}\\
    \hline
    $n$-bit Addition & Complete \\
    $n$-bit Less-than Comparison & Partial \\
    Beaver Triple Generation & None \\
    \hline
  \end{tabular}
  \bigskip
  \caption{MPC frameworks and instantiated functions used in our evaluation. Each framework describes an approach for evaluating a binary circuit in MPC; we instantiate both frameworks with three different functions (represented as circuits) to construct secure protocols. }
  \label{tbl:protocols}
\end{table}

\paragraph{GMW (Goldreich-Micali-Widgerson).}
The GMW protocol~\cite{goldreich2019play, goldreich2009foundations} is an MPC protocol for evaluating circuits. Each party holds \emph{additive secret shares} of each wire's value; to secret share a bit, a party selects a random bit to represent one share, and computes the other share by XORing the random bit with the secret bit.
The parties evaluate addition gates (and other linear operations) by adding the shares they hold.
The parties evaluate multiplication gates by using 1-out-of-4 oblivious transfer: the first party generates a random bit for its share of the output, and acts as the sender in OT; the second party acts as the receiver in OT and receives its share of the output. We implemented a compiler to generate \langname programs that evaluate Bristol fashion circuits using GMW.

\paragraph{MPC with Beaver Triples.}
Another MPC protocol for evaluating circuits also uses additive shares to hold wire values, but uses \emph{Beaver triples}~\cite{beaver1992efficient} to evaluate multiplication gates (instead of OT). A Beaver triple (or \emph{multiplication triple}) is a set of secret shares of numbers $a$, $b$, and $c$ such that $a \cdot b = c$. To use a Beaver triple, $P_1$ must have one share of each of $a$, $b$, and $c$, and $P_2$ must have the other share. Since the values are secret shared, they appear random to each party---but the correlation between the three numbers can be used to perform efficient multiplication between secret-shared numbers (described later). One Beaver triple must be consumed for each multiplication, but they can be generated in a separate offline phase.

Given shares of a Beaver triple, the parties perform two multiplication-by-constant and broadcast operations, and derive one share each of the product. Our implementation of this approach employees a third-party ``dealer'' who generates the triples; a pre-processing (offline) phase can also be used to generate the triples. We implemented a similar compiler to generate \langname programs leveraging this approach for circuit specifications.

\paragraph{Functions evaluated.}
We evaluate three functions, which we express as binary circuits for two parties and compile to \langname programs using the approach described above. All three functions expect input from both the honest and corrupt party. The first, an $n$-bit addition circuit, is vacuously secure---its ideal functionality leaks all of the honest party's inputs, because the corrupt party can subtract their inputs from the $n$-bit result to reconstruct the honest party's inputs. The second, an $n$-bit less-than comparison circuit, only leaks a single bit of information---which of the two inputs is larger. For $n=1$, the less-than protocol is also vacuously secure, but it leaks only partial information about the honest party's inputs for $n>1$. The third, an $n$-bit Beaver triple generator, leaks no information about the honest party's input, since its output to each party is a single secret share of the Beaver triple.

\begin{figure*}
  \centering
  \newcommand{\gsize}{.45\textwidth}
\begin{tabular}{c| c c}
    \hline\hline
  & \textbf{OT-Based (GMW)} & \textbf{Beaver Triple-Based}\\
    \hline\hline
  \rotatebox{90}{\phantom{hellohello}$n$-bit addition}
  & \includegraphics[width=\gsize]{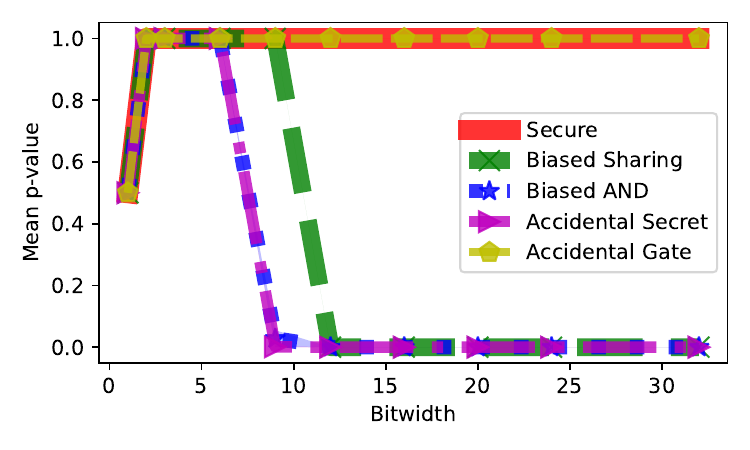}
                 & \includegraphics[width=\gsize]{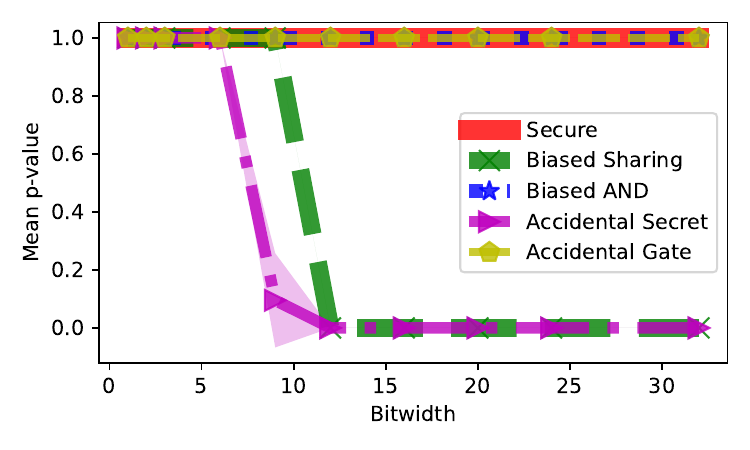} \\
    \hline
  \rotatebox{90}{\phantom{hel}$n$-bit less-than comparison}
  & \includegraphics[width=\gsize]{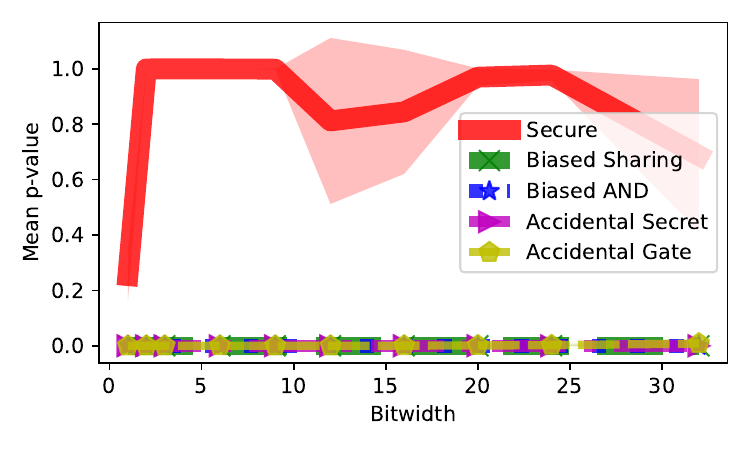}
                 & \includegraphics[width=\gsize]{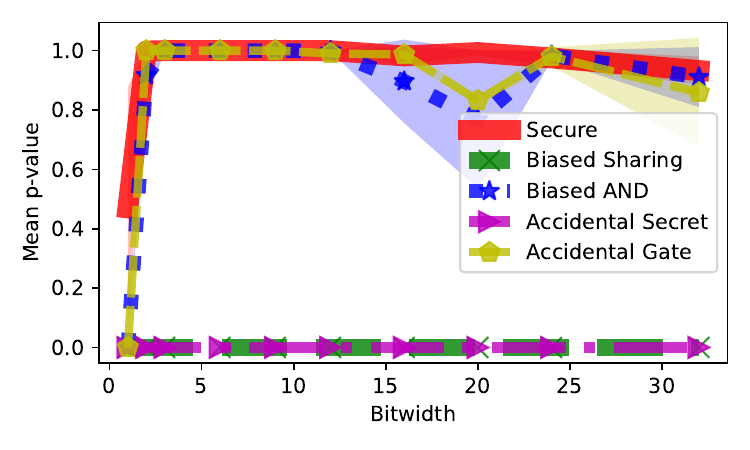} \\
    \hline
  \rotatebox{90}{\phantom{h}$n$-bit Beaver triple generation}
  & \includegraphics[width=\gsize]{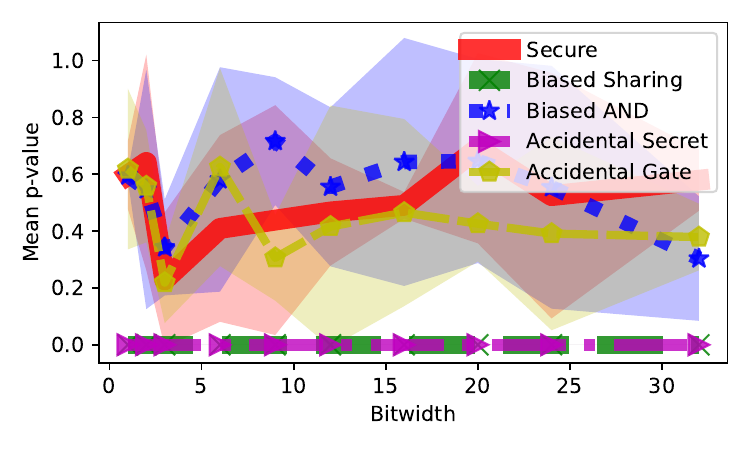}
                 & \includegraphics[width=\gsize]{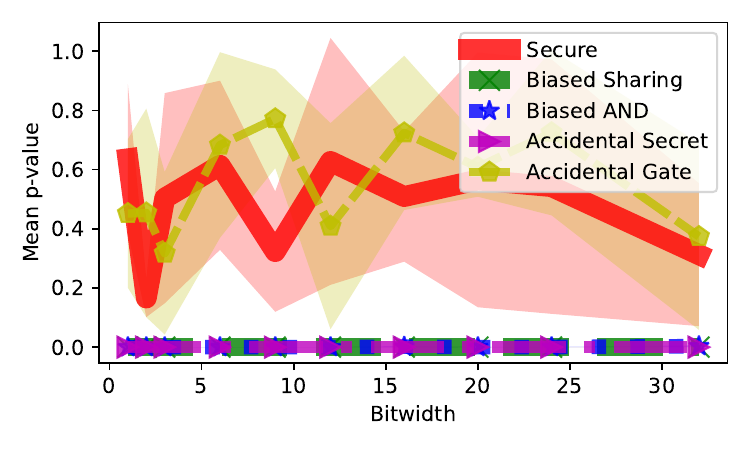} \\
    \hline
    \hline
\end{tabular}
\caption{Experimental results: ability to detect security bugs. We set the number of iterations to 128 and the training set size to 1024. Results for other settings appear in Appendix~\ref{sec:addit-exper-results}.}
\label{fig:results_security}
\end{figure*}

\begin{figure*}
  \centering
  \newcommand{\gsize}{.45\textwidth}
\begin{tabular}{c| c c}
    \hline\hline
  & \textbf{OT-Based (GMW)} & \textbf{Beaver Triple-Based}\\
    \hline\hline
  \rotatebox{90}{\phantom{hellohello}$n$-bit addition}
  & \includegraphics[width=\gsize]{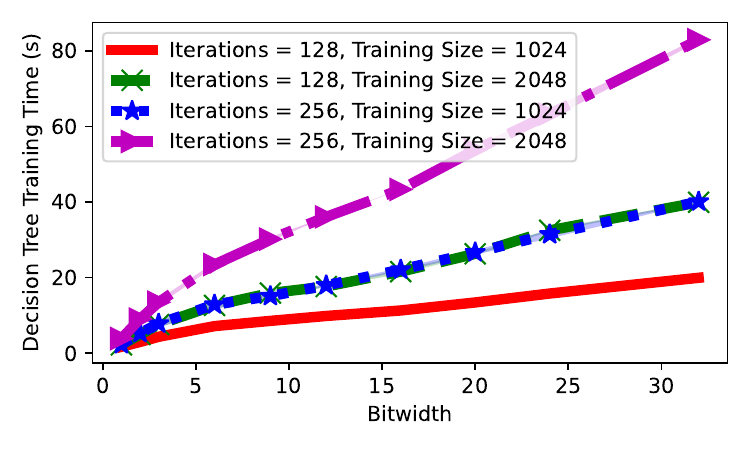}
                 & \includegraphics[width=\gsize]{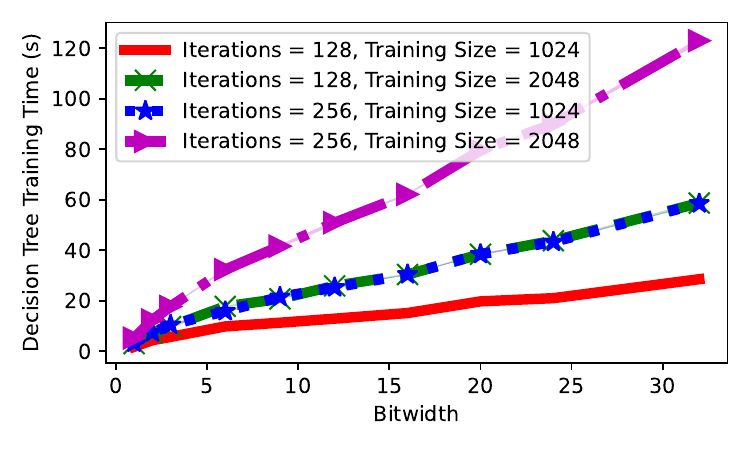} \\
    \hline
  \rotatebox{90}{\phantom{hel}$n$-bit less-than comparison}
  & \includegraphics[width=\gsize]{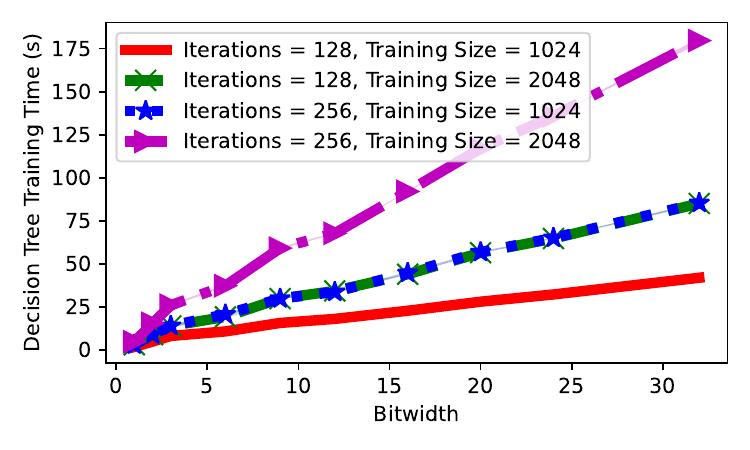}
                 & \includegraphics[width=\gsize]{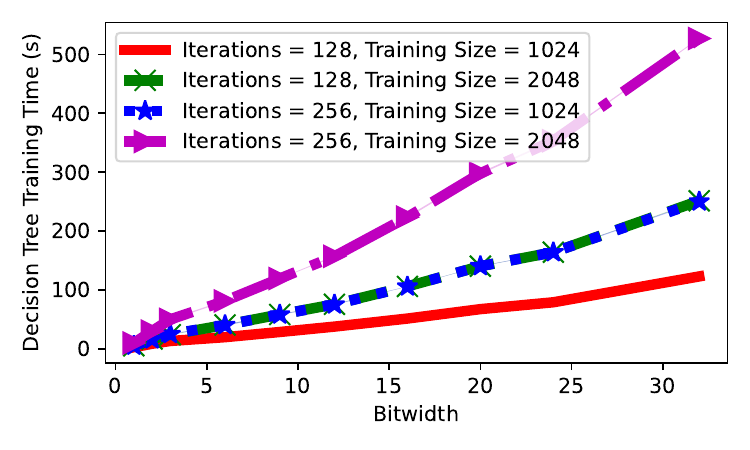} \\
    \hline
  \rotatebox{90}{\phantom{h}$n$-bit Beaver triple generation}
  & \includegraphics[width=\gsize]{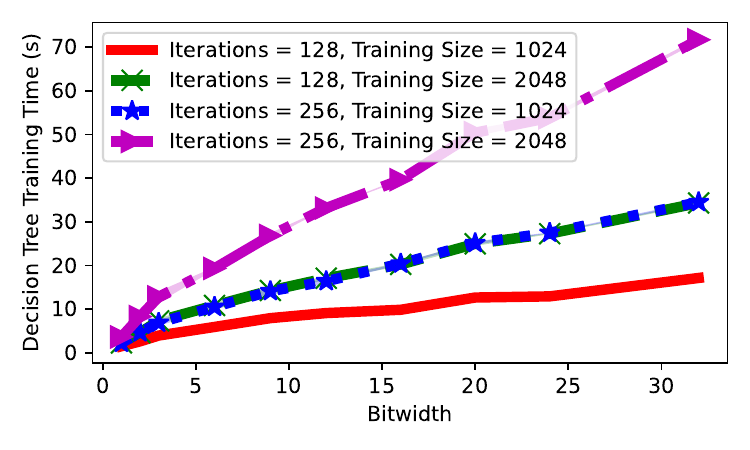}
                 & \includegraphics[width=\gsize]{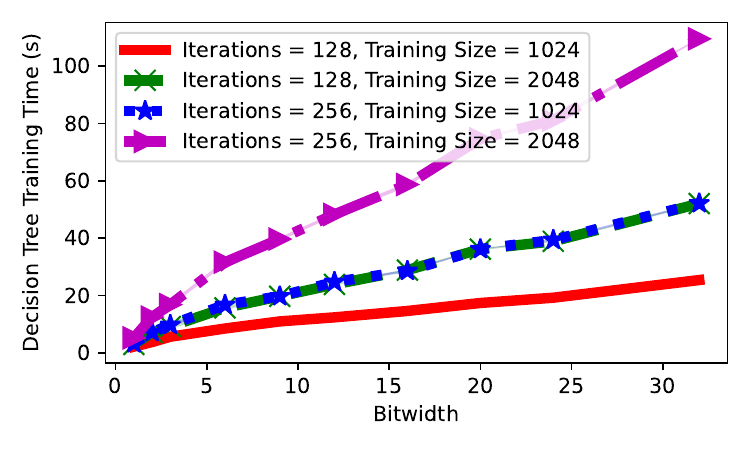} \\
    \hline
    \hline
\end{tabular}
\caption{Experimental results: decision tree training time.}
\label{fig:results_training_time}
\end{figure*}

\subsubsection{Experiment Setup}
\label{sec:e2-experiment-setup}

We run \toolname on the unmodified protocol in order to establish the protocol's security
and show that \toolname doesn't flag secure protocols as insecure.
Then we mutate the protocol in several ways to introduce security bugs and test \toolname's ability to detect these bugs. Specifically, our mutations are:
\begin{itemize}
\item \textbf{Biased sharing}: uses biased randomness to construct secret shares of honest party inputs.
\item \textbf{Biased AND}: uses biased randomness as part of the evaluation of multiplication gates.
\item \textbf{Accidental secret}: sometimes, at random, accidentally sends a secret held by the honest party  to the corrupt party.
\item \textbf{Accidental gate}: sometimes, at random, accidentally sends an honest party's result from the evaluation of a multiplication gate to the corrupt party.
\end{itemize}
The severity of each mutation can be tuned by adjusting the bias in the related randomness. In the case of \textbf{accidental secret} and \textbf{accidental gate}, this affects the probability of accidentally revealing a value; in the others, it directly adjust the bias in the randomness.

\subsubsection{Results}
\label{sec:e2_results}

The results appear in Figures~\ref{fig:results_security} and~\ref{fig:results_training_time}. For the results in Figure~\ref{fig:results_security}, we set the number of iterations to 128 and the training set size to 1024. The results are similar for other settings; additional experimental results appear in Appendix~\ref{sec:addit-exper-results}.

\paragraph{Detecting security bugs.}
\toolname is able to detect most security bugs in our evaluation, but interpreting its results requires care. \toolname detects \textbf{biased sharing} bugs in nearly all cases: the associated $p$-values are universally close to 0 in all six cases. The one exception is the $n$-bit adder for small $n$, where associated $p$ values are close to 1. This is likely due to the \emph{vacuous security} of the $n$-bit adder: for small bitwidths, the decision trees are able to reconstruct the honest party's inputs using only information from the ideal functionality, rendering \emph{any} protocol secure. For larger bitwidths, the decision trees are no longer able to perform this reconstruction, and the protocols are classified as insecure.

\toolname detects \textbf{biased AND} bugs in most of the GMW-based protocols, but struggles to do so in the case of Beaver-triple-based protocols. Due to the use of multiplication triples, taking advantage of the leakage from biased AND gates in this case is much more difficult than in the GMW case. These results suggest that \toolname will likely have difficulty detecting bugs that require significant ``backwards computation'' to leverage the leaked information.

\toolname detects \textbf{accidental secret} bugs in the same cases as biased sharing bugs, for the same reasons.

\toolname detects \textbf{accidental gate} bugs in the $n$-bit-less-than comparison only in the OT-based framework.
In the Beaver-triple framework, we hypothesize that the decision trees are not capable of the ``backwards computation''
required to leverage the leaked information

In other cases, the function's design means that AND gate outputs do not provide additional leakage beyond the ideal functionality,
and therefore do not compromise security.
\toolname ignores \textbf{biased AND} and \textbf{accidental gate} bugs in the case of $n$-bit Beaver triple generation
because the result of each AND gate is also one of the protocol's outputs---so a bug which reveals it does not leak any new information.
\toolname detects \textbf{accidental gate} bugs for the $n$-bit Beaver triple generation function when the Beaver-triple-based protocol is used, because biasing the randomness for AND gates reveals information about the Beaver triples themselves, from which it is possible to deduce the honest party's shares of the inputs.

\paragraph{Scalability.}
The primary computational cost of \toolname is the time taken to train the decision trees. We record the total training time for each combination of MPC framework and function, and plot the results for several settings in Figure~\ref{fig:results_training_time}. For the settings we evaluated, training time grows linearly with the number of iterations, the training data size, and the bitwidth of the target function (i.e. the size of the circuit being tested). Even for the largest bitwidths we considered, training takes only seconds to minutes.

\subsection{Discussion}

The results of these experiments suggest that \toolname scales linearly with protocol size and test power, and is capable of detecting bugs in both randomly-generated and real-world protocols. The test typically runs in just seconds to minutes. The results of our experiment on protocol mutations show that \toolname is effective for detecting important classes of security bugs, but that care must be taken in interpreting the results.

For security bugs that directly reveal sensitive information (e.g. \textbf{biased sharing} bugs) \toolname is particularly effective. Our results suggest that \toolname is capable of detecting these accidental disclosures even at low test powers and in the context of large protocols.

\toolname is less effective for security bugs that require significant ``backwards computation'' to leverage the leaked information (e.g. \textbf{accidental gate} bugs), since decision trees are not well-suited to solving this problem. A more sophisticated model---such as a deep neural network---may be able to perform this kind of computation, but would take significantly longer to train.

\section{Related Work}

\subsection{Formal-methods for cryptography}
Although automated language-based verification of MPC security is an open question,
a variety of useful work approaches the subject from different directions.

Haagh et al.~\cite{haagh2018computer} % Computer-aided proofs for multiparty computation with active security
use EasyCrypt to manually show the security of GMW.
Their definitions generalize to protocols with efficiently invertible functionalities.
They model MPC as a non-interference property: Honest secrets are non-interfering on the corrupt views conditioned on the corrupt outputs.
Their approach requires manual proof in EasyCrypt.

Barthe et al.~\cite{barthe2019probabilistic} % https://arxiv.org/abs/1907.10708 A Probabilistic Separation Logic
give a framework for probabilistic reasoning and show that it can be used to prove security of additive secret sharing.
However, the framework is not amenable to automation.

Gancher et al.~\cite{gancher2023core} % https://dl.acm.org/doi/pdf/10.1145/3571223
give a powerful language system that can check if one protocol is a valid simulator of another,
and show that it works on existing MPC protocols like GMW.
The user of the system needs to write a second protocol
representing the composition of the ideal-world functionality and an explicit $\mathtt{SIM}$,
and then build a high-level proof of the equivalence in the provided equational logic.

The \textsc{Wysteria} language~\cite{rastogi2014wysteria} % https://www.ieee-security.org/TC/SP2014/papers/Wysteria_c_AProgrammingLanguageforGeneric,Mixed-ModeMultipartyComputations.pdf
gives fine-grained integration of MPC as a language primitive, to help a larger program use it efficiently.
Their system assumes (at least one) secure implementation of MPC on circuits.

$\lambda_\textbf{obliv}$~\cite{darais2019language} %
gives a type system that ensures uniform and independent randomness of key values throughout a program.
Their system is automated, and could be adapted to ensure perfect noninterference between honest secrets and corrupt views,
but does not support the kind of \emph{conditional} independence required for most MPC protocols.

\textsc{Owl}~\cite{gancher2023owl} % https://eprint.iacr.org/2023/473.pdf
gives a type system guaranteeing non-interference with an (extensible) selection of cryptographic operations as language primitives.
OWL is designed for secure protocols that reveal nothing to the adversary, and does not support verification of MPC security.

Fournet et al.~\cite{fournet2011information} % https://www.microsoft.com/en-us/research/wp-content/uploads/2017/01/information-flow-types-for-homomorphic-encryptions-ccs11.pdf
also use a type system to ensure safe usage of cryptographic primitives,
but they specifically work with homomorphic encryption schemes and the various measures by which such schemes may be correct.
Like \textsc{Owl}, their type system assumes a secure implementation of the encryption scheme and does not support verification of MPC security.

\subsection{PPLs for Inference}

Several PPLs directly support efficient inference and some forms of independence testing.
A Sum Product Network (SPN)~\cite{poon2011sum} is a directed-graph representation of a multi-variate distribution;
SPPL~\cite{saad2021sppl} is a PPL that performs inference using SPNs.
SPNs are compact, allow efficient inference, and can represent values symbolically as needed,
so testing for conditional independence is practical.
However, such a language cannot express any existing MPC protocols, and likely cannot express any \textit{useful} protocols,
because if the language allowed any value to be used more than once, it would result in graphs that violate key invariants of SPNs.

The \textsc{Dice} language~\cite{holtzen2020scaling} analogously translates programs into binary decision diagrams (BDDs).
BDDs can represent, and do inference on, distributions that aren't representable as SPNs,
but since the graph \textit{structure} of the BDD of a \textsc{Dice} program depends on the actual \textit{values} on which one is conditioning,
a \textsc{Dice}-based test for the kind of conditional independence needed for MPC would scale exponentially with the input/output width of the protocol.

\textsc{Lilac}~\cite{li2023lilac} % https://arxiv.org/abs/2304.01339
is a program logic for reasoning about probability distributions and PPLs.
It more expressive than the original probabilistic separation logic,
but does not support automation.

\subsection{Probabilistic testing}

Ding et al.~\cite{ding2018detecting} % https://arxiv.org/pdf/1805.10277.pdf
and more recently Zhang et al.~\cite{zhang2020testing} % https://dl.acm.org/doi/10.1145/3428233
use large samples of program executions to test for violations of differential privacy,
showing that such techniques can work on probabilistic hyper-properties.
Since differential privacy is only concerned with program inputs and outputs,
the dimensionality of the problem is somewhat smaller than MPC.

Our work builds directly on that of Chalupka et al.~\cite{chalupka2018fast}
who show that batches of ML classifiers can be used to check conditional independence.
Specifically, if decision trees (D-Trees) trained on random variables $A$ and $C$
do better at predicting the value of a random variable $B$ than D-Trees trained only on $C$,
then $A$ and $B$ are not independent conditioned on $C$.
The test is statistical; many D-Trees are trained and any apparent advantage of the better-informed ones must be assessed for statistical significance.
In their work the data is assumed to be permuted sub-samples of some finite pool of samples from a population,
but in our adaptation we use fresh program traces for everything.

\section{Future Work}

Some further work is needed to turn \toolname into a deploy-able tool;
our experiments so far are not sufficient to calculate sensible default settings,
and we believe some changes to the way the data is ingested and handled by the tool could improve efficiency.

There are also variations on the core concept of \toolname that we believe are worth exploring.
Decision trees are not the only candidate for the underlying ML model, and our evaluation suggests that they are incapable of performing some of the ``backwards computation'' needed to demonstrate insecurity;
a deeper model (\eg{} a neural network) may be more capable of detecting these kinds of security bugs.
Other cryptographic properties besides MPC can likely be framed in terms of conditional independence;
and so would be amenable to similar tests.
Finally, because end-to-end static analysis of MPC protocols has proven so difficult,
we believe it may be practical to augment a stochastic tool like \toolname with static analysis capabilities
that would allow it to more efficiently detect insecurity, turning it into a "white-box" test.

\section{Conclusions}

We present \toolname,
a property-based test that checks (a small over-approximation of) MPC security
using a probabilistic test of conditional independence.
At its heart, \toolname trains batches of decision trees with and without access to grey-box data from many evaluations of the target protocol
to see if access to that message data improves their ability to predict the secret inputs.
While this approach has theoretical limitations, we show that in practice it is able to detect various mistakes in various protocols,
making it a useful tool for sanity-checking one's work during development.
We also show that, in practice, the compute-time required for \toolname scales linearly with the power of the test requested,
and usefully-powerful tests take just a few minutes.

\section*{Acknowledgments}

This material is based upon work supported by the National Science Foundation under Grant No. 2238442 and by the Cold Regions Research and Engineering Laboratory (ERDC-CRREL)
under Contract No. W913E521C0003. Any opinions, findings and conclusions or recommendations expressed in this material are those of the author(s) and do not necessarily reflect the views of the National Science Foundation or the Cold Regions Research and Engineering Laboratory.

\bibliographystyle{plain}
\bibliography{refs}

\appendices

\section{Additional Experimental Results}
\label{sec:addit-exper-results}

This Appendix contains additional experimental results excluded from the main body of the paper:
\begin{itemize}
\item Figure~\ref{fig:extra1} presents additional results for $n$-bit addition.
\item Figure~\ref{fig:extra2} presents additional results for $n$-bit less-than comparison.
\item Figure~\ref{fig:extra3} presents additional results for $n$-bit Beaver triple generation.
\end{itemize}

\input{full_figures.tex}

\section{Example \langname program}\label{sec:example-cho}

The following \langname uses oblivious-transfer based GMW to evaluate a circuit representing a "less than" function
between the two parties' inputs, interpreted as 2-bit binary numbers.
It is slightly shorter than the ones we used for our experiments;
to provide the closest possible comparison between different implementations in our experiments,
the \langname protocols we used included some "no-op" statements.

\begin{small}
\begin{verbatim}
-- two_bit_less_than.cho
MACRO secret_share(P1(x), P2()) AS
  s1 = FLIP @P1
  s2 = x + s1
  SEND s2 TO P2
ENDMACRO
MACRO and_gmw(P1(x2, y2), P2(x1, y1)) AS
  out2 = FLIP @P2
  g1_s2_00 = out2 + ((x1 + 0) ^ (y1 + 0))
  g1_s2_01 = out2 + ((x1 + 0) ^ (y1 + 1))
  g1_s2_10 = out2 + ((x1 + 1) ^ (y1 + 0))
  g1_s2_11 = out2 + ((x1 + 1) ^ (y1 + 1))
  out1 = OBLIVIOUSLY [[g1_s2_00,
                       g1_s2_01]?y2,
                      [g1_s2_10,
                       g1_s2_11]?y2
                     ]?x2 FOR P1
ENDMACRO
MACRO reveal(P1(x1), P2(x2)) AS
  SEND x1 TO P2
  SEND x2 TO P1
  y = x1 + x2
ENDMACRO
-- Read secrets
x0 = SECRET @P1
x1 = SECRET @P1
y2 = SECRET @P2
y3 = SECRET @P2
-- Set up shares
DO secret_share(P1(x0), P2()) GET(x0_1=s1,
                                  x0_2=s2)
DO secret_share(P1(x1), P2()) GET(x1_1=s1,
                                  x1_2=s2)
DO secret_share(P2(y2), P1()) GET(y2_1=s2,
                                  y2_2=s1)
DO secret_share(P2(y3), P1()) GET(y3_1=s2,
                                  y3_2=s1)
-- Circuit evaluation
g3_1 = ~y2_1
g3_2 = y2_2
DO and_gmw(P1(x0_1, g3_1), P2(x0_2, g3_2)
          ) GET(g4_1=out1, g4_2=out2)
g5_1 = ~x0_1
g5_2 = x0_2
DO and_gmw(P1(y2_1, g5_1), P2(y2_2, g5_2)
          ) GET(g6_1=out1, g6_2=out2)
g7_1 = ~y3_1
g7_2 = y3_2
g8_1 = ~g6_1
g8_2 = g6_2
DO and_gmw(P1(g7_1, g8_1), P2(g7_2, g8_2)
          ) GET(g9_1=out1, g9_2=out2)
DO and_gmw(P1(x1_1, g9_1), P2(x1_2, g9_2)
          ) GET(g10_1=out1, g10_2=out2)
DO and_gmw(P1(g4_1, g10_1), P2(g4_2, g10_2)
          ) GET(g11_1=out1, g11_2=out2)
g12_1 = g4_1 + g10_1
g12_2 = g4_2 + g10_2
g13_1 = g11_1 + g12_1
g13_2 = g11_2 + g12_2
-- Reveal output
DO reveal(P1(g13_1), P2(g13_2)) GET(r0=y)
OUTPUT r0
\end{verbatim}
\end{small}

\end{document}

%% file: full_figures.tex
% These are full figures for the appendix

\begin{figure*}
  \centering
  \newcommand{\gsize}{.45\textwidth}
\begin{tabular}{c| c c}
    \hline\hline
  & \textbf{OT-Based (GMW)} & \textbf{Beaver Triple-Based}\\
    \hline\hline
  \rotatebox{90}{\phantom{helloh}$i = 128, n = 1024$}
  & \includegraphics[width=\gsize]{graphs/security_adder_gmw_128_1024.pdf}
                 & \includegraphics[width=\gsize]{graphs/security_adder_beaver_128_1024.pdf} \\
    \hline
  \rotatebox{90}{\phantom{helloh}$i = 128, n = 2048$}
  & \includegraphics[width=\gsize]{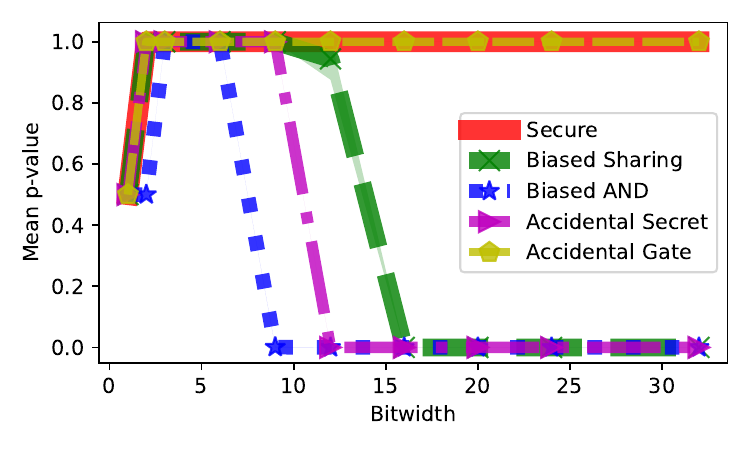}
                 & \includegraphics[width=\gsize]{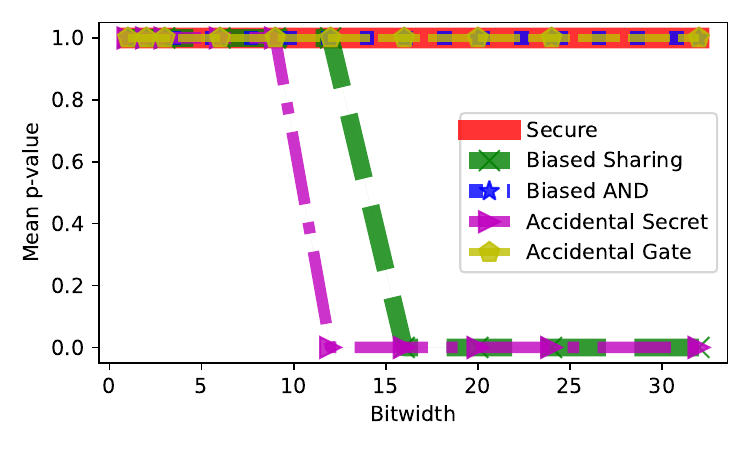} \\
    \hline
  \rotatebox{90}{\phantom{helloh}$i = 256, n = 1024$}
  & \includegraphics[width=\gsize]{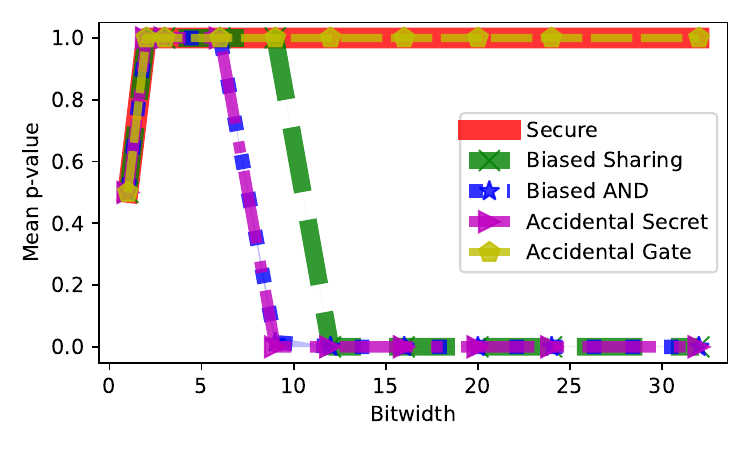}
                 & \includegraphics[width=\gsize]{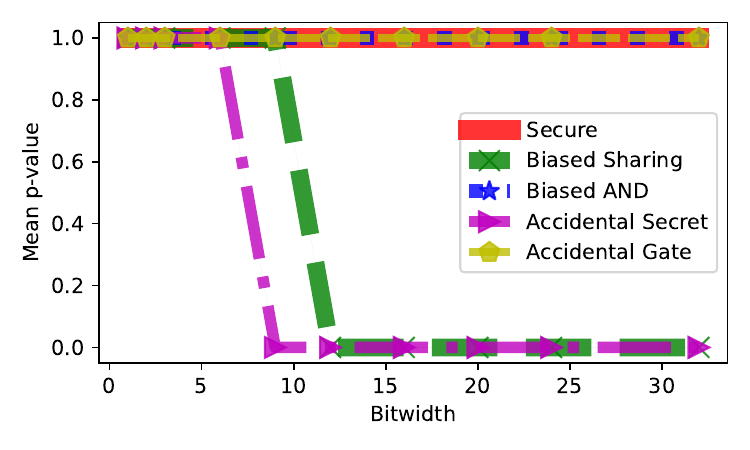} \\
    \hline
  \rotatebox{90}{\phantom{helloh}$i = 256, n = 2048$}
  & \includegraphics[width=\gsize]{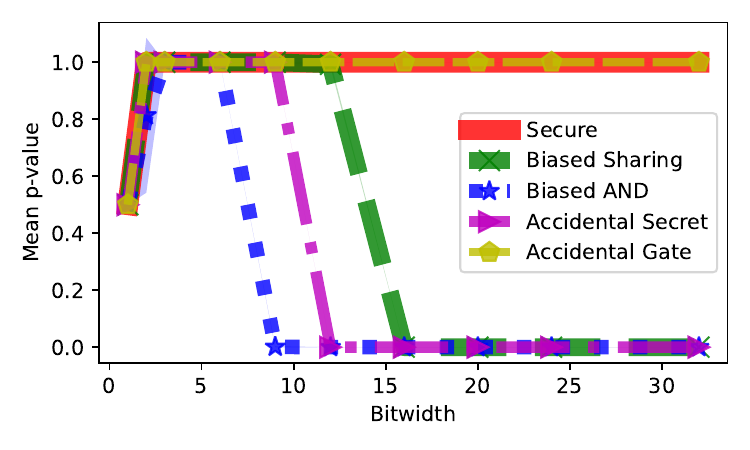}
                 & \includegraphics[width=\gsize]{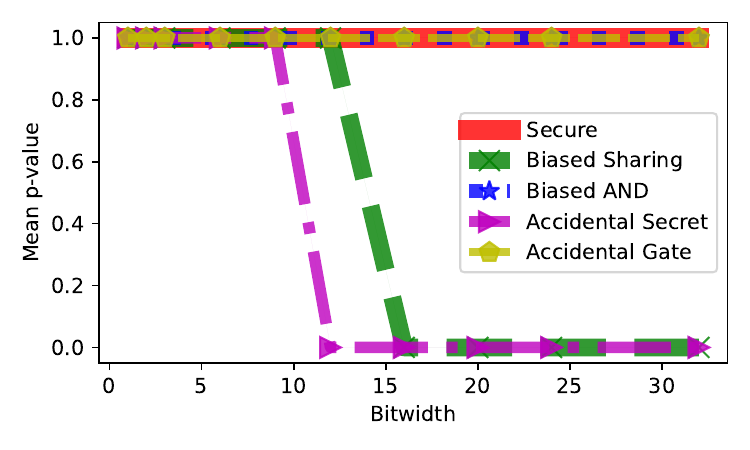} \\
    \hline
    \hline
\end{tabular}
\caption{Experimental results, $n$-bit addition.}
\label{fig:extra1}
\end{figure*}

\begin{figure*}
  \centering
  \newcommand{\gsize}{.45\textwidth}
\begin{tabular}{c| c c}
    \hline\hline
  & \textbf{OT-Based (GMW)} & \textbf{Beaver Triple-Based}\\
    \hline\hline
  \rotatebox{90}{\phantom{helloh}$i = 128, n = 1024$}
  & \includegraphics[width=\gsize]{graphs/security_less_than_gmw_128_1024.pdf}
                 & \includegraphics[width=\gsize]{graphs/security_less_than_beaver_128_1024.pdf} \\
    \hline
  \rotatebox{90}{\phantom{helloh}$i = 128, n = 2048$}
  & \includegraphics[width=\gsize]{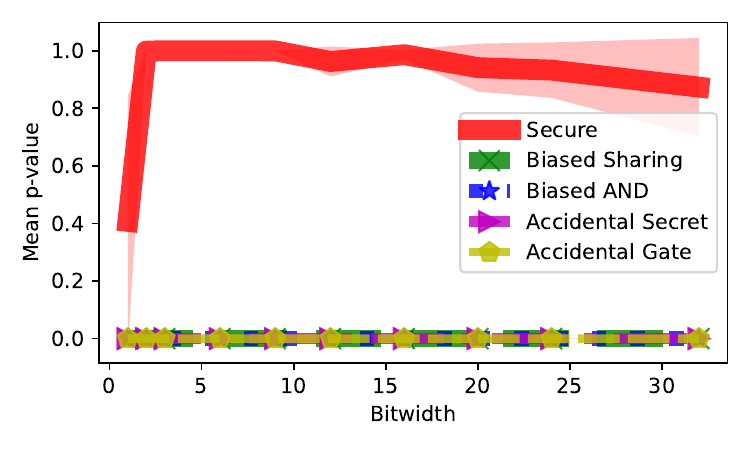}
                 & \includegraphics[width=\gsize]{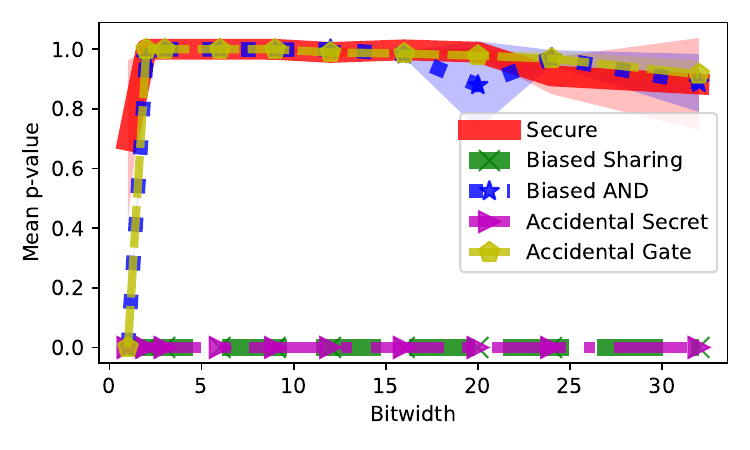} \\
    \hline
  \rotatebox{90}{\phantom{helloh}$i = 256, n = 1024$}
  & \includegraphics[width=\gsize]{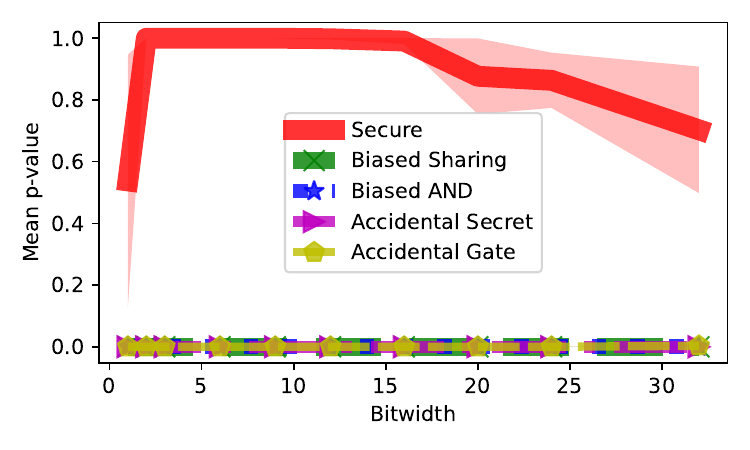}
                 & \includegraphics[width=\gsize]{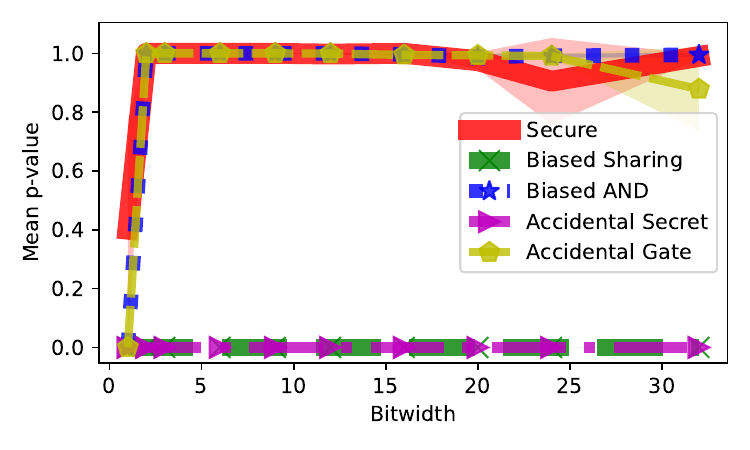} \\
    \hline
  \rotatebox{90}{\phantom{helloh}$i = 256, n = 2048$}
  & \includegraphics[width=\gsize]{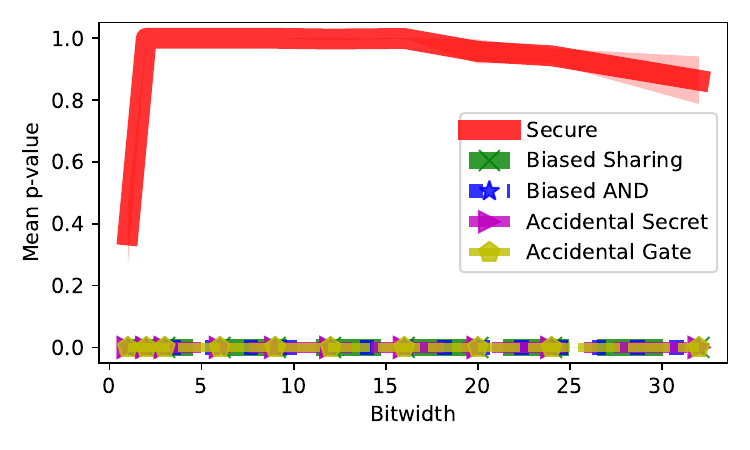}
                 & \includegraphics[width=\gsize]{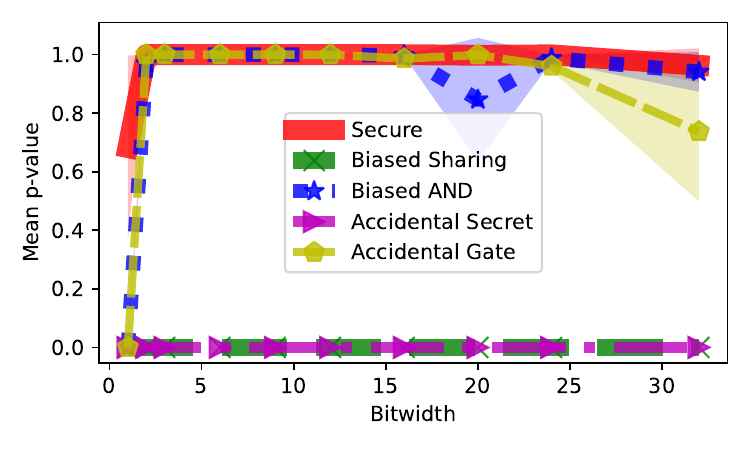} \\
    \hline
    \hline
\end{tabular}
\caption{Experimental results, $n$-bit less-than comparison.}
\label{fig:extra2}
\end{figure*}

\begin{figure*}
  \centering
  \newcommand{\gsize}{.45\textwidth}
\begin{tabular}{c| c c}
    \hline\hline
  & \textbf{OT-Based (GMW)} & \textbf{Beaver Triple-Based}\\
    \hline\hline
  \rotatebox{90}{\phantom{helloh}$i = 128, n = 1024$}
  & \includegraphics[width=\gsize]{graphs/security_beaver_triple_gen_gmw_128_1024.pdf}
                 & \includegraphics[width=\gsize]{graphs/security_beaver_triple_gen_beaver_128_1024.pdf} \\
    \hline
  \rotatebox{90}{\phantom{helloh}$i = 128, n = 2048$}
  & \includegraphics[width=\gsize]{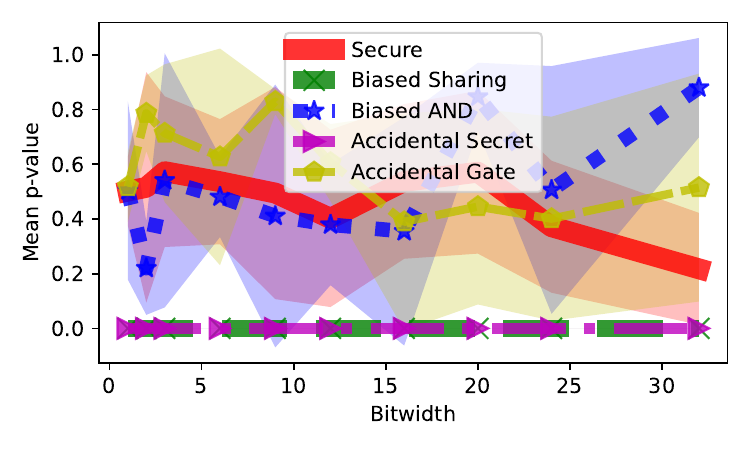}
                 & \includegraphics[width=\gsize]{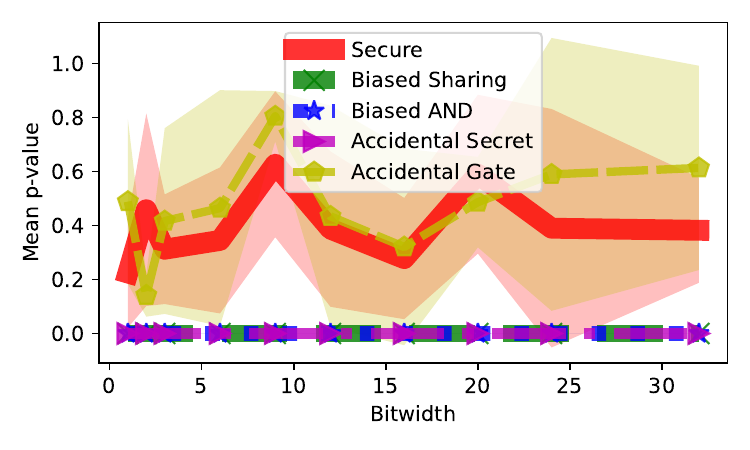} \\
    \hline
  \rotatebox{90}{\phantom{helloh}$i = 256, n = 1024$}
  & \includegraphics[width=\gsize]{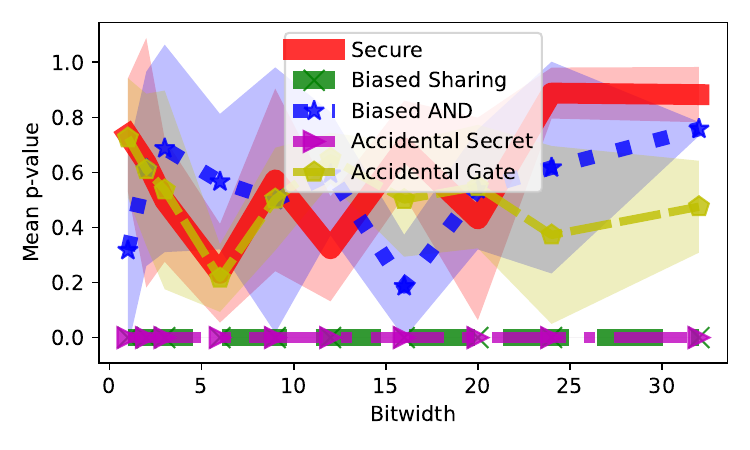}
                 & \includegraphics[width=\gsize]{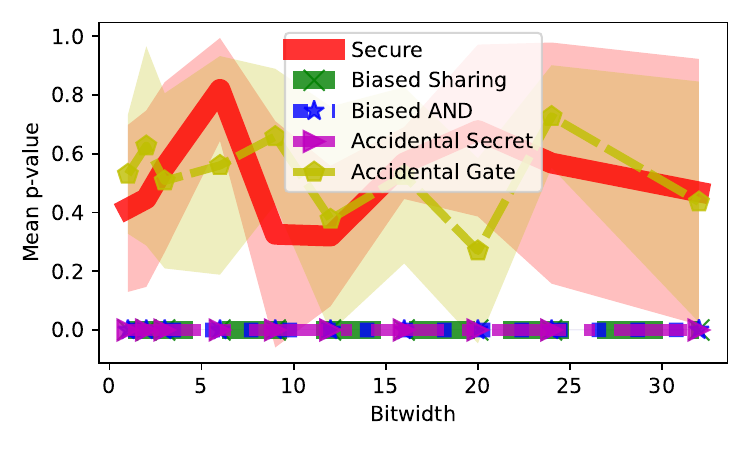} \\
    \hline
  \rotatebox{90}{\phantom{helloh}$i = 256, n = 2048$}
  & \includegraphics[width=\gsize]{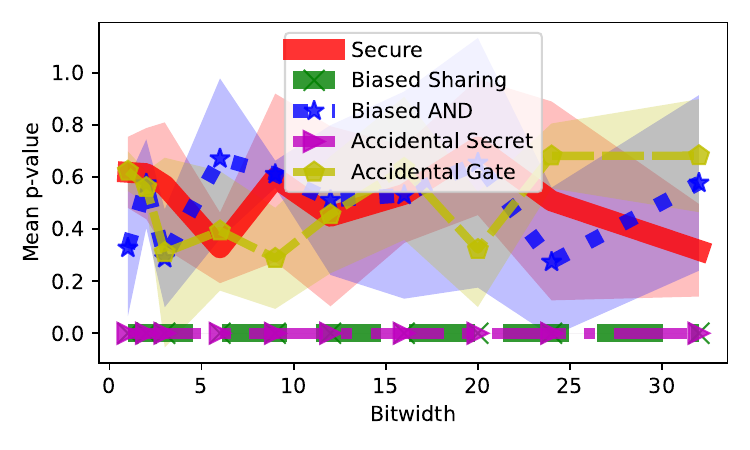}
                 & \includegraphics[width=\gsize]{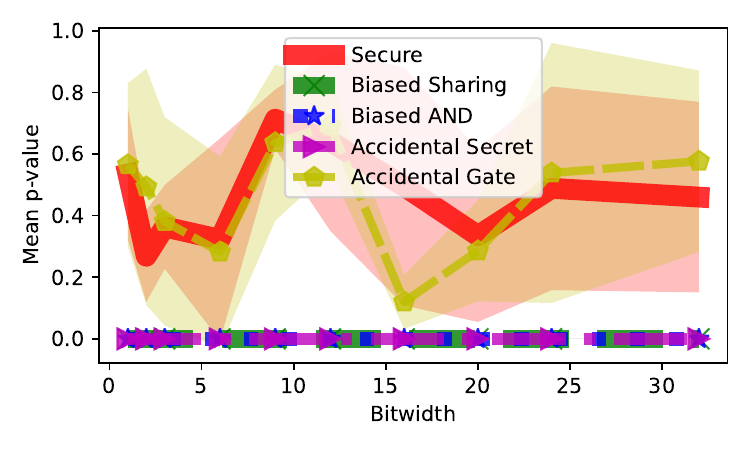} \\
    \hline
    \hline
\end{tabular}
\caption{Experimental results, $n$-bit Beaver triple generation.}
\label{fig:extra3}
\end{figure*}